# A quantum phase switch between a single solid-state spin and a photon

Shuo Sun,[1] Hyochul Kim,[1] Glenn S. Solomon,[2] and Edo Waks[1,a]

[1]Department of Electrical and Computer Engineering, Institute for Research in Electronics and Applied Physics, and Joint Quantum Institute, University of Maryland, College Park, Maryland 20742, USA

[2]Joint Quantum Institute, National Institute of Standards and Technology, and University of Maryland, Gaithersburg, Maryland 20899, USA

**Abstract**

Strong interactions between single spins and photons are essential for quantum networks and distributed quantum computation. They provide the necessary interface for entanglement distribution, non-destructive quantum measurements, and strong photon-photon interactions. Achieving spin-photon interactions in a solid-state device could enable compact chip-integrated quantum circuits operating at gigahertz bandwidths. Many theoretical works have suggested using spins embedded in nanophotonic structures to attain this high-speed interface. These proposals exploit strong light-matter interactions to implement a quantum switch, where the spin flips the state of the photon and a photon flips the spin-state. However, such a switch has not yet been realized using a solid-state spin system. Here, we report an experimental realization of a spin-photon quantum switch using a single solid-state spin embedded in a nanophotonic cavity. We show that the spin-state strongly modulates the cavity reflection coefficient, which conditionally flips the polarization state of a reflected photon on picosecond timescales. We also demonstrate the complementary effect where a single photon reflected from the cavity coherently rotates the spin. These strong spin-photon interactions open up a promising direction for solid-state implementations of high-speed quantum networks and on-chip quantum information processors using nanophotonic devices.

---
[a] Email: edowaks@umd.edu.

The ability to store and transmit quantum information plays a central role in virtually all quantum information processing applications. Single spins serve as pristine quantum memories while photons are ideal carriers of quantum information. Strong interactions between these two systems provide the necessary interface for developing future quantum networks[1] and distributed quantum computers[2]. They also enable a broad range of critical quantum information functionalities such as entanglement distribution[3,4], non-destructive quantum measurements[5-7], and strong photon-photon interactions[8-10]. Realizing spin-photon interactions in a solid-state device is particularly desirable because it opens up the possibility for chip-integrated quantum circuits that support gigahertz bandwidth operation[11].

Many theoretical works have suggested using nanophotonic cavities or waveguides to attain a high-bandwidth interface between spins and photons in a solid-state device[12-18]. These proposals exploit strong light-matter interactions to create a quantum switch that applies a spin-dependent quantum phase or amplitude modulation to a photon. Quantum switches have been demonstrated using a single trapped atom in a cavity[6,7,9], but they have not yet been realized in a solid-state spin system.

Quantum dots provide one promising approach to realize quantum switches in a solid-state device. When coupled to nanophotonic structures, they exhibit strong light-matter interactions[19] that can interface stationary solid-state quantum bits (qubits) with photons, as previously demonstrated using optically excited quantum dot states[20]. But such excited states have short lifetimes that make them impractical for quantum information processing applications. The spin of a single electron or hole trapped in a charged quantum dot provides a much better qubit system that supports microsecond coherence times[21,22] and picosecond timescale single-qubit operations[23,24]. Charged quantum dots are also optically active and can directly emit spin-entangled photons[25-28]. In addition, the integration of charged quantum dots with cavities has recently experienced significant progress. Several works have demonstrated deterministic loading of a spin in a quantum dot coupled to a nanophotonic cavity[29-31], coherent control of the loaded spin[32], and large Kerr rotations[33]. However, the demonstration of a coherent quantum switch, the essential building block for quantum information processing, remains an outstanding challenge.

In this article we report an experimental demonstration of a quantum phase switch using a single solid-state spin embedded in a nanophotonic cavity. We implement this switch using a spin trapped in a charged quantum dot that is strongly coupled to a photonic crystal defect cavity. We show that the switch applies a spin-dependent phase shift on a reflected photon that rotates its polarization state. We also demonstrate the complementary effect where a single reflected photon applies a π phase shift to one of the spin-states and thereby coherently rotates the spin. These results demonstrate that the quantum switch retains phase coherence, an essential requirement for quantum information applications. We demonstrate switching of photon wavepackets as short as 63 ps, which corresponds to a three orders of magnitude increase in bandwidth over atom-based quantum switches. Our work represents a critical step towards interconnecting multiple solid-state spins using photons for implementing quantum networks and distributed quantum information protocols.

**OPERATING PRINCIPLE**

Figure 1a illustrates the energy level diagram for a negatively charged quantum dot containing a single electron[34]. Positively charged dots would have a similar level structure[34]. The states of the dot include two ground states, corresponding to the two electron spin orientations, and two excited trion states that optically couple to the ground states via four optical transitions denoted as $\sigma_1$ - $\sigma_4$.

We implement a quantum phase switch based on the original proposal of Duan and Kimble[8], which utilizes an atomic qubit coupled to a single-sided cavity. In the ideal implementation, transition $\sigma_1$ is resonant with the cavity mode while the remaining transitions are decoupled due to large detuning. Thus, the quantum dot resonantly couples to the cavity only when it is in the spin-up state, inducing a spin-dependent reflection coefficient. For a monochromatic photon that is resonant with the $\sigma_1$ transition, the reflection coefficients for the spin-down and spin-up states are $r_\downarrow = -(2\alpha - 1)$ and $r_\uparrow = 1 - \frac{2\alpha}{1+C}$ respectively[12,35-37] (see Supplementary Section S1 for a complete derivation of the reflection coefficients). Here, $C = 2g^2/\kappa\gamma$ is the atomic cooperativity, which is a function of the coupling strength $g$, the total cavity energy decay rate $\kappa$, and the dipole decay rate $\gamma$, and $\alpha = \kappa_{ex}/\kappa$ is the interference contrast, where $\kappa_{ex}$ is the cavity energy decay rate to the reflected mode. For an ideal single-sided cavity $\alpha = 1$, but realistic cavities may suffer from intra-cavity losses and decay to parasitic photonic channels which serve to degrade the interference. When $C > 1$ and $\alpha > 0.5$, $r_\uparrow$ and $r_\downarrow$ have opposite signs. Thus, the spin-state conditionally shifts the phase of a reflected photon by $\pi$, implementing a quantum phase operation. We attain an ideal phase switch in the limit of large cooperativity ($C \gg 1$) and a perfect single-sided cavity ($\alpha = 1$) where the reflection coefficient becomes $r_\downarrow = -1$ and $r_\uparrow = 1$.

The quantum phase switch allows one qubit to conditionally switch the quantum state of the other qubit. We consider the case where the polarization state of the photon encodes quantum information. Since photonic crystal cavities have a single mode with a well-defined polarization, we can express the state of a photon incident on the cavity in the basis states $|x\rangle$ and $|y\rangle$, which denote the polarization states oriented orthogonal and parallel to the cavity mode respectively. The quantum state of a right-circularly polarized incident photon can be written as $|\psi_i\rangle = |x\rangle + i|y\rangle$. Upon reflection the state becomes $|\psi_f\rangle = |x\rangle + ir_{\uparrow(\downarrow)}|y\rangle$, which remains unchanged if the quantum dot is in the spin-up state, but becomes left-circularly polarized for spin-down. Similarly, if the spin is prepared in the state $|\psi_i\rangle = |\uparrow\rangle + |\downarrow\rangle$, then after a $y$-polarized photon reflects from the cavity the spin-state transforms to $|\psi_f\rangle = |\uparrow\rangle - |\downarrow\rangle$, which creates a spin-flip corresponding to a rotation of $\pi$ along the equator of the Bloch sphere.

**SPIN-DEPENDENT CAVITY REFLECTIVITY**

Figure 1b shows a scanning electron microscope image of the fabricated photonic crystal cavity. The Methods section describes the device design and fabrication procedure. We initially characterize the device using micro-photoluminescence measurements (see Supplementary Section S2). From these measurements, we identify a single charged quantum dot that is strongly coupled to the cavity mode and is red-detuned by 67 GHz.

To demonstrate that the spin can flip the state of the photon, we use the polarization interferometry setup shown in Fig. 1c (see Methods). We excite the cavity with right-circularly polarized light, and measure the reflected signal along either the left-circularly or right-circularly polarized component. Figure 2a shows both the cross-polarized (red diamonds) and co-polarized reflection spectrum (blue circles) when the quantum dot is detuned from the cavity so that the two systems are decoupled. By fitting the reflection spectrum to a Lorentzian lineshape (blue and red solid lines; see Supplementary Section S3), we determine the cavity energy decay rate to be $\kappa/2\pi = 35.9 \pm 0.7 \text{ GHz}$, and the interference contrast to be $\alpha = 0.81 \pm 0.01$.

We next apply a magnetic field of 6.6 T that tunes transition $\sigma_1$ onto cavity resonance via a Zeeman shift. We excite the quantum dot with a narrowband tunable laser in order to optically pump the spin-state[38,39]. We first tune the optical pumping laser to transition $\sigma_4$ to prepare the quantum dot in the spin-up state. The blue circles in Fig. 2b show the cross-polarized reflection spectrum with the optical pumping laser, which exhibits a vacuum Rabi splitting. When we turn off the pumping laser, we observe a reduced contrast due to random spin fluctuations (red diamonds). Co-polarized measurements show the expected complementary behavior (see Supplementary Section S4). In contrast, when we optically pump transition $\sigma_2$ to initialize the quantum dot to the spin-down state, we observe a spectrum that closely resembles a bare cavity (Fig. 2c). This spin-dependent reflection spectrum is one of the essential properties of the phase switch. We note that we were not able to perform a co-polarized measurement when optically pumping transition $\sigma_2$ because the pump laser is too close to the probe and generates large background.

We numerically fit the measured spectra to a theoretical model (blue solid line; see Supplementary Section S3). From the fit to Fig. 2b we determine the coupling strength between $\sigma_1$ and the cavity to be $g_1/2\pi = 10.2 \pm 0.1 \, \text{GHz}$, and the dipole decay rate of $\sigma_1$ to be $\gamma_1/2\pi = 2.9 \pm 0.4 \, \text{GHz}$, corresponding to a cooperativity of $C = 2$. This cooperativity is sufficiently large to enable quantum phase switching operation. The coupling strength satisfies the condition $g_1 > \kappa/4$, which indicates that we are operating in the strong coupling regime[19]. We fit the data in Fig. 2c to a model that considers a mixture of two spin-states. From this fit we determine the occupation probability of the spin-down state to be $74\% \pm 8\%$. We attribute this imperfect spin initialization to spin frustration caused by pumping the $\sigma_3$ transition, which is close to resonance with $\sigma_2$. In this case frustration is enhanced because the Raman scattering process is resonant with the cavity.

To quantitatively characterize the performance of the switch, we define the switching fidelity conditioned on collecting a reflected photon as $F_{\uparrow(\downarrow)} = |\langle \psi_{d,\uparrow(\downarrow)} | \psi_{f,\uparrow(\downarrow)} \rangle|^2$, where $\psi_{d,\uparrow(\downarrow)}$ denotes the photon state for an ideal switch given by $\psi_{d,\uparrow} = \frac{1}{\sqrt{2}}(|x\rangle + i|y\rangle)$ and $\psi_{d,\downarrow} = \frac{1}{\sqrt{2}}(|x\rangle - i|y\rangle)$, and $\psi_{f,\uparrow(\downarrow)}$ is the actual state of the reflected photon. In Supplementary Section S5 we provide a detailed analysis of the switching fidelity that account for finite cooperativity, intra-cavity losses, coupling and collection losses from the cavity, and imperfect transverse spatial mode-matching. From this analysis we calculate a switching fidelity of 0.88 and 0.95 for the spin-up and spin-down states respectively.

**COHERENT CONTROL OF CAVITY REFLECTIVITY**

To demonstrate control of a reflected photon using a coherently prepared spin state, we use all-optical coherent control (see Methods). A narrowband continuous-wave laser tuned to transition σ4 performs spin initialization, while circularly polarized picosecond optical pulses generate an effective spin rotation[23,24]. We perform spin-rotations using 6 ps rotation pulses with center frequencies detuned by 520 GHz from the cavity resonance (equal to 15 cavity linewidth). In order to rotate the spin over the Bloch sphere, we utilize the Ramsey interferometry setup illustrated in Fig. 3a, which generates a pair of rotation pulses separated by a time delay $\tau$. A third laser pulse probes the cavity reflectivity a time $\Delta t$ after the second rotation pulse. We attenuate this laser so that a single pulse contains an average of 0.12 photons coupled to the cavity, in order to ensure a low probability of two-photon events. We set the power of the continuous-wave optical pumping laser to 30 nW. At this power we measure a spin initialization time of $1.27 \pm 0.09$ ns (see Supplementary Section S6), which is slow compared to $\tau$ and $\Delta t$, but fast compared to the repetition time of the experiment (13 ns).

Figure 3b shows the reflected probe intensity as a function of rotation pulse power $P$ and delay $\tau$, where we set $\Delta t$ to 140 ps. We observe Ramsey oscillations in the reflected probe intensity as a function of both $P$ and $\tau$. Figure 3c plots the emission intensity of the quantum dot at transition $\sigma_2$ for the same measurement, which provides a second readout of the spin-state. We observe the same Ramsey oscillation pattern in the quantum dot emission signal, confirming that the reflection modulation shown in Fig. 3b is induced by coherent spin manipulation. Figure 3d shows the numerically calculated value for the population of the spin-down state for comparison (see Supplementary Section S7), which exhibit good agreement with experiments.

In Fig. 3e we plot the reflected probe intensity over a larger time range of 1 ns. We fix the power of each rotation pulse to 40 µW which corresponds to a π/2-rotation. From the decay of the fringe visibility, we calculate a $T_2^*$ of $0.94 \pm 0.02$ ns. This coherence time is limited by inhomogeneous broadening due to a slowly fluctuating nuclear spin background[21], along with decoherence due to continuous optical pumping during the rotation pulse sequence. We could reduce these effects by turning off the pump laser during the measurement process and using a nuclear field locking[21] or spin echo technique[22], which has been shown to improve the coherence to up to 2.6 µs.

To characterize the fidelity of the spin-state preparation, we tune the probe laser across the cavity resonance while setting $P$ and $\tau$ to the conditions indicated by the circles in Fig. 3b. The resulting cavity spectra are plotted in Fig. 4. In Fig. 4a the two pulses arrive in-phase with the Larmor precession of the spin, and the quantum dot rotates to the spin-down state. The cavity spectrum (blue circles) is thus similar to the bare cavity Lorentzian lineshape. Figure 4b shows the case where the two rotation pulses arrive out-of-phase and the quantum dot rotates back to the spin-up state. The cavity now (blue circles) exhibits a strongly coupled spectrum. We also plot the measured spectrum when $\Delta t = 13$ ns (red diamonds) for comparison. At this condition the spin is re-initialized to the spin-up state in both cases.

We numerically fit the data to a model that is similar to the one used for Fig. 2, but also accounts for the 7 GHz bandwidth of the probe pulse (see Supplementary Section S3). From the fit we determine the spin preparation fidelity in these two conditions to be $0.70 \pm 0.04$ and $0.74 \pm 0.05$. The imperfect population transfer could be caused by a number of factors such as re-initialization of the spin by optical pumping during the interval between the rotation and the probe pulse, and decoherence due to power induced trion dephasing induced by the rotation pulses[24,40].

**CONTROLLING A SPIN WITH A PHOTON**

The previous measurements demonstrate that the spin-state of the quantum dot induces a conditional phase shift on the photon. A quantum phase switch would also exhibit the complementary effect, where reflection of a single photon rotates the spin-state. To demonstrate this phase shift, we use the experimental configuration shown in Fig. 5a. We again perform a Ramsey interference measurement but we inject a weak laser pulse that serves as the control field before the second rotation pulse (see Methods). We generate the control pulse in the same way as the probe pulse in the previous measurement, with pulse duration of 63 ps. When a control photon couples to the cavity, it imposes a phase shift on the spin-down state, which shifts the phase of the Ramsey fringes.

We perform spin readout by monitoring the emission at the $\sigma_2$ frequency. The blue circles in

Fig. 5b show the occupation probability of the spin-down state conditioned on the detection of a control photon, as a function of delay between the two rotation pulses. These data are obtained by performing a two photon correlation measurement. The blue solid line is a numerical fit to a sinusoidal function. We compare this curve to the occupation probability of the spin-down state when we block the control field (black squares with black line as a numerical fit). The interference fringe conditioned on detecting a single control photon is shifted in phase by $(1.09\pm0.09)\pi$ radians relative to the case where there is no control photon, demonstrating that a single control photon applies a large phase shift to the spin. We attribute the degraded visibility of the Ramsey fringe conditioned on a control photon to finite cooperativity, intrinsic cavity losses, and occasional two-photon incidence events.

We can tune the phase shift imparted on the spin by a control photon by introducing a frequency detuning between the control field and transition $\sigma_1$, which enables us to apply arbitrary controlled phase shifts. Figure 5c shows the same measurement for a blue detuned control field. The conditioned data (blue circles) show a $(0.59\pm0.05)\pi$ radian phase shift, which corresponds to a detuning of 7.3 GHz (see Supplementary Section S8). We also plot the occupation probability of the spin-down state in the presence of the control field but without conditioning on the detection of the control photon (red diamonds in Fig. 5b and 5c). These curves are very similar to the case where the control field is blocked, which indicates that the average number of control photons per pulse coupled to the cavity is much smaller than one.

**DISCUSSION**

We have demonstrated a solid-state spin-photon quantum phase switch, which is the fundamental building block for numerous applications including entanglement distribution[3,4], non-destructive photon measurements[5-7], and strong photon-photon interactions[8-10]. The large light-matter coupling strength of our solid-state devices enables a phase switch operating at unprecedented bandwidths, where the spin can switch photon wavepackets as short as 63 ps. Perhaps the most intriguing aspect of our quantum switch is that it monolithically combines spins with strongly interacting nanophotonic structures on a single semiconductor chip, which may have many beneficial properties for future integration and scalability.

The performance of the device could be improved in several ways. Smaller mode-volume cavity designs could enable higher switching fidelities by improving the cooperativity[41]. Using a delta-doping layers[22,24,25] or active charge stabilization[32] could further improve the spin-state preparation fidelity. Our results can also be directly applied in waveguide integrated devices that are more conducive to on-chip integration and can exhibit similar strong light-matter interactions[42]. In such on-chip implementations, waveguide losses create additional challenges by degrading the cavity $Q$, which would reduce the cooperativity. Waveguide-coupled devices would therefore require much higher bare cavity $Q$ to ensure that the light remains on the chip. A number of works have reported a bare cavity $Q$ in GaAs photonic crystal cavities exceeding 250,000[43,44], which could potentially enable both efficient on-chip coupling and high cooperativity. Employing regulated quantum dot growth techniques[45,46] in conjunction with local frequency tuning[47] could further open up the possibility to integrate multiple switches on a single semiconductor chip. Ultimately, solid-state quantum phase switches could play an important role in scaling semiconductor quantum devices to more complex quantum systems for applications in quantum networking, computation, and simulation.

**Methods**

**Device design and fabrication.** The initial wafer for device fabrication was composed of a 160 nm thick GaAs membrane with a single layer of InAs quantum dots at its center (density of 10-50/$\mu m^2$). A fraction of quantum dots in the sample were naturally charged due to residual doping background. We used weak white light illumination to stabilize the extra electron confined in the dot. The membrane layer was grown on top of a 900 nm thick $Al_{0.78}Ga_{0.22}As$ sacrificial layer. A distributed Bragg reflector composed of 10 layers of GaAs and AlAs was grown below the sacrificial layer and acted as a high reflectivity mirror, creating a one-sided cavity. Photonic crystal structures were defined using electron-beam lithography, followed by inductively coupled plasma dry etching and selective wet etching of the sacrificial AlGaAs layer. The cavity design was based on a three-hole defect in a triangular photonic crystal[48], with a lattice constant of 240 nm and a hole radius of 72 nm.

**Measurement setup**. The sample was mounted in a closed-cycle cryostat and cooled down to 3.6 K. A superconducting magnet was used to apply magnetic fields up to 9.2 T along the in-plane direction (Voigt configuration). The cavity axis was rotated approximately 45 degree relative to the magnetic field. Sample excitation and collection was performed with a confocal microscope using an objective lens with numerical aperture of 0.68. The optimal coupling efficiency for this configuration was measured to be 1% by measuring the Stark shift of the quantum dot under cavity-resonant excitation[47]. The polarization of the incident light and collected signals were set by a quarter-wave plate and a polarizer, as illustrated in Fig. 1c. The collected signal was focused into a single-mode fiber that spatially filtered out spurious surface reflection. The spectrum of the signal was measured using a grating spectrometer with a resolution of 7 GHz.

**Spin dependent cavity reflectivity measurement.** Data in Fig. 2 were measured using a narrowband tunable external cavity diode laser with linewidth below 300 kHz. The laser was sent through an intensity stabilizer before coupling to the excitation port. A second narrow linewidth (< 100 kHz) external cavity diode laser was used to resonantly pump one of the quantum dot transitions. Each data was obtained by numerically fitting the measured reflection with a Gaussian function, where the frequency and reflected intensity were obtained from the fit. This technique enabled us to monitor the laser frequency to much higher precision than the spectrometer resolution.

**Ramsey interferometry measurement**. The rotation and probe pulses were generated using two time-synchronized Ti:sapphire lasers with a repetition rate of 76 MHz. The lasers were synchronized by piezo feedback in the rotation laser cavity, which locked its clock frequency to the probe laser with an accuracy of 100 fs. The delay between the rotation and the probe pulse $\Delta t$ was controlled electronically by a phase-lock loop in the synchronization electronics, and measured by a single-photon avalanche photodiode with 30 ps resolution. The probe pulse duration, initially 2 ps, was filtered to 63 ps using a spectrometer grating, and then sent to an intensity stabilizer. The rotation and probe pulses were sent to the excitation port shown in Fig. 1c, and the polarizations were set the same as the measurements for Fig. 2b and 2c.

**Photon induced spin phase switch measurement.** The signal from the control pulse and the quantum dot emission were collected using the same objective lens, and then separated into two optical paths with a 50/50 beamsplitter. A 9-GHz-bandwidth Fabry-Perot filter was used in one of the optically paths to spectrally select only the photons emitted at the frequency of the $\sigma_2$ transition. A polarizer and a grating spectrometer were used in the other optical path to select photons tuned to the control field frequency and polarized parallel to the cavity-axis. To obtain the blue circles in Fig. 4b and 4c, we performed a two-photon coincidence measurement between the control photon and the quantum dot emission at the $\sigma_2$ transition using two Single Photon Counting Modules (SPCMs) and a PicoHarp 300 time correlated single photon counting system. At each delay time $\tau$ between the two rotation pulses, we count the two-photon coincidence events $C(\tau)$ and calculate the conditioned probability $P(\tau)$ as

$$P(\tau) = \frac{C(\tau)}{\max_{\tau}\{C(\tau)\} + \min_{\tau}\{C(\tau)\}}.$$

To obtain the red diamonds and blue squares, we only measure the quantum dot emission intensity at the $\sigma_2$, denoted $I(\tau)$, using a spectrometer CCD and calculate the probability as $P(\tau) = \frac{I(\tau)}{\max_{\tau}\{I(\tau)\} + \min_{\tau}\{I(\tau)\}}$. The control field was turned on and off to obtain the red diamonds and black squares respectively.

**Acknowledgments:** The authors would like to acknowledge support from the Army Research Office Multidisciplinary University Research Initiative on Hybrid quantum interactions (grant number W911NF09104), the DARPA QUINESS program (grant number W31P4Q1410003), the Physics Frontier Center at the Joint Quantum Institute, and the Office of Naval Research Applied Electromagnetics center. E. W. would like to acknowledge support from the National Science Foundation Faculty Early Career Development (CAREER) award (grant number ECCS. 0846494).

**Author Contributions:** S.S., H.K. and E.W. conceived and designed the experiment. S.S. and E.W. prepared the manuscript and carried out the theoretical analysis. S.S. carried out the measurement and analyzed the data. H.K. contributed to the measurement and sample design. G.S.S. provided samples grown by molecular beam epitaxy.

**Competing Financial Interests:** The authors declare no competing financial interests.

# Figure Legends

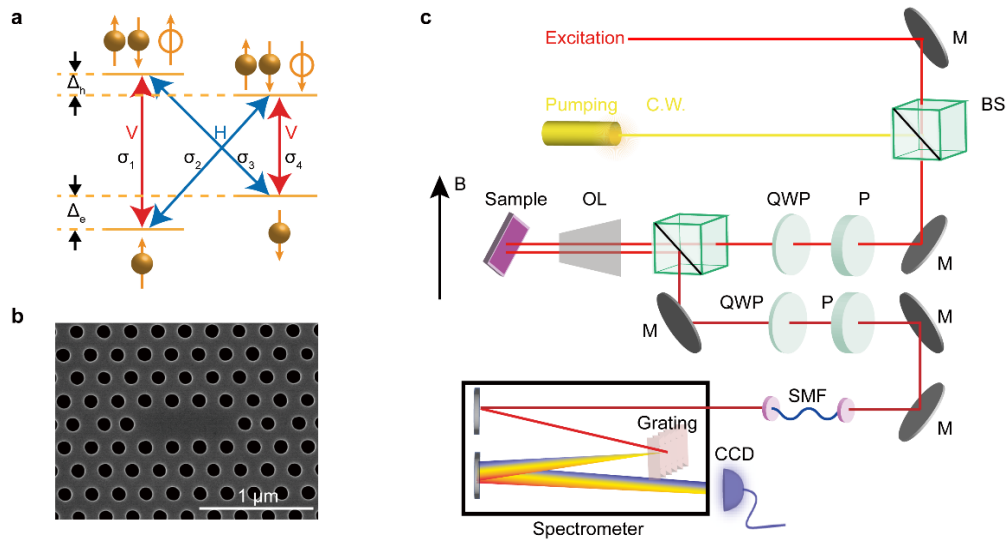

**Figure 1 | Device and experimental setup. a,** Energy level structure of a negatively charged quantum dot under a magnetic field. The level structure is composed of two spin ground states and two excited trion states, with four allowed optical transitions labeled σ₁ to σ₄. The vertical and cross transitions couple to orthogonal linear polarizations of light, denoted *V* and *H* respectively. **b,** Scanning electron microscope image of the fabricated device. **c,** Measurement setup. OL, objective lens; QWP, quarter wave plate; P, polarizer; BS, beam splitter; M, mirror; SMF, single mode fiber; CCD, charged coupled device.

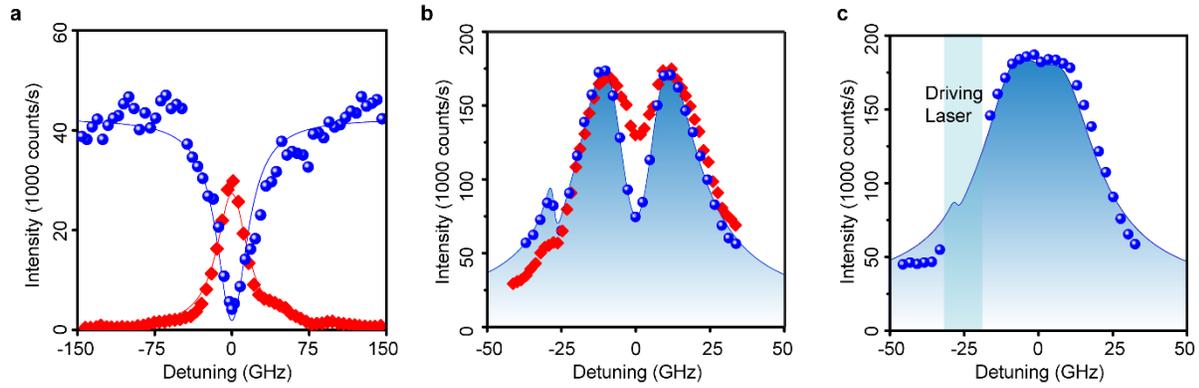

**Figure 2 | Spin dependent cavity reflectivity. a,** Co-polarized (blue circles) and cross-polarized (red diamonds) cavity reflection spectrum with no magnetic field. Blue and red solid lines show the calculated spectrum. **b,** Cavity reflection spectrum at 6.6 T magnetic field with (blue circles) and without (red diamonds) an optical pumping laser resonant with transition $\sigma_4$. The blue solid line shows the calculated spectrum for the case where the optical pumping laser is turned on. With the pumping laser, we observe a suppression of the cavity response at the $\sigma_1$ resonance due to strong coupling. We also observe a Fano-resonant lineshape at -27 GHz detuning, corresponding to the coupling between $\sigma_2$ and the cavity mode. **c,** Cavity reflection spectrum when the pump laser is resonant with transition $\sigma_2$. The blue circles show the measured spectrum, and the solid line shows calculated spectrum. The center wavelength is 927.48 nm for all panels.

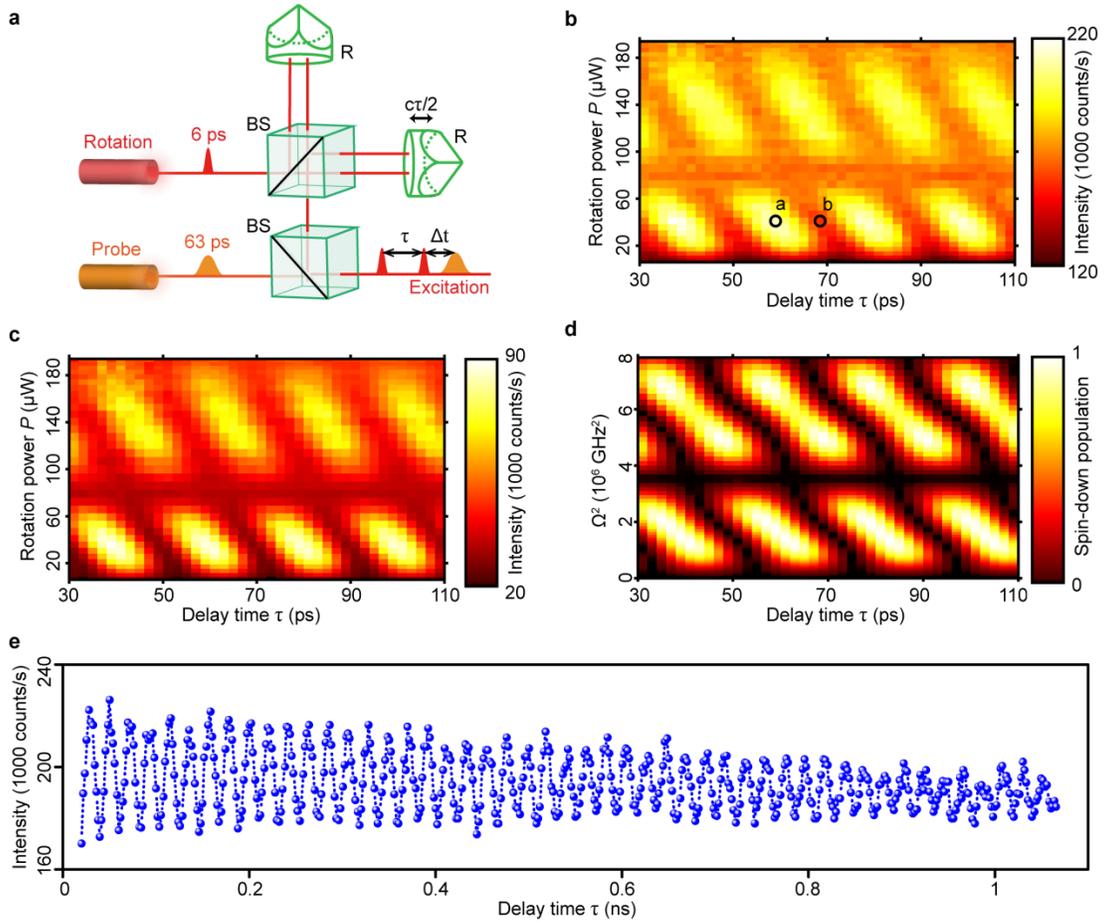

**Figure 3 | Ramsey interferometry measurements. a,** Experimental setup for generating the Ramsey pulse sequence. The delay time $\tau$ between the two rotation pulses is controlled by a movable retro-reflector mounted on a computer-controlled translation stage. BS, beam splitter; R, retro-reflector. **b,** Reflected probe intensity as a function of rotation pulse power $P$ and the delay time $\tau$. **c,** Intensity of the quantum dot emission at $\sigma_2$ transition frequency as a function of rotation pulse power $P$ and the delay time $\tau$. **d,** Calculated spin-down state population as a function of peak rotation pulse power and the delay time $\tau$. We express the rotation pulse as a classical time-varying Rabi frequency with a Gaussian pulse shape and peak power $\Omega^2$. **e,** Reflected probe intensity as a function of delay time $\tau$.

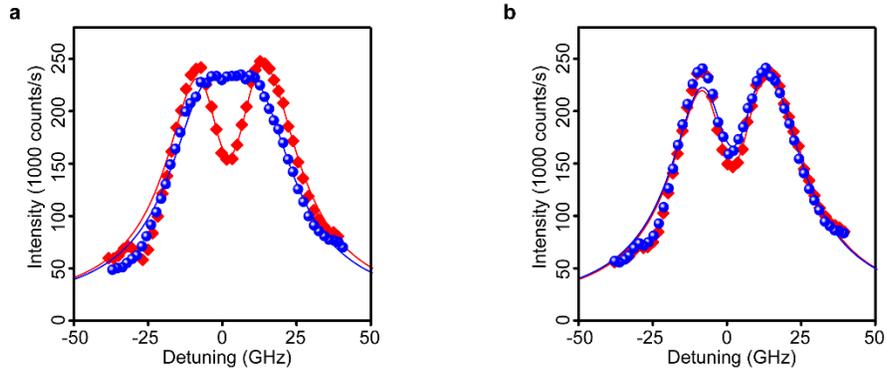

**Figure 4 | Time resolved cavity reflection spectrum. a-b,** Reflected probe intensity as a function of probe detuning at the rotation condition indicated by (**a**) point "a" and (**b**) point "b" in Fig. 3b. Blue circles, $\Delta t = 140\,\text{ps}$; red diamonds, $\Delta t = 13\,\text{ns}$. Solid lines are the calculated spectra. The center wavelength is 927.48 nm for both spectra.

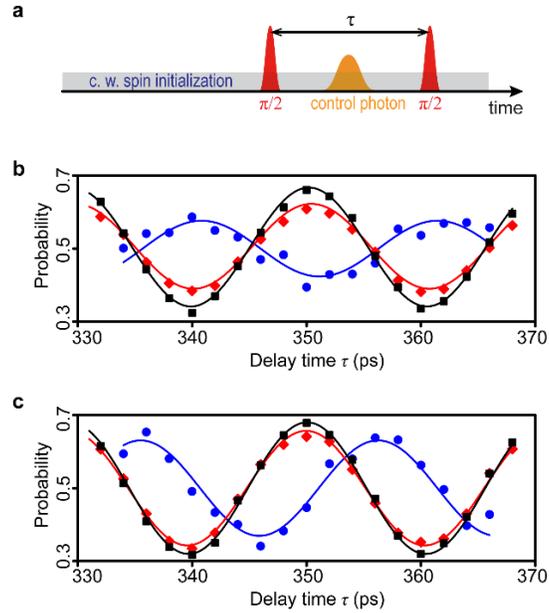

**Figure 5 | Photon induced spin phase switch. a,** Pulse timing diagram showing the relative time delays between the rotation pulses and the control field. **b,** Occupation probability of the spin-down state as a function of the delay time $\tau$, in the absence of control pulse (black squares), conditioned on detecting a reflected control photon polarized along the cavity axis (blue circles), and in the presence of the control field but not conditioning on the detection of a control photon (red diamonds). The control field is resonant with the $\sigma_1$ quantum dot transition. **c,** Same as **b**, except that the control field is blue detuned from the $\sigma_1$ quantum dot transition.